\documentstyle[psfig,twocolumn,aps,floats]{revtex}
\begin{document}
\wideabs{
\title{Low temperature value of the 
upper critical field $H_{\lowercase{c}2}(0)$ 
 of isotropic single  band metals}
\author{S.V.\ Shulga$^{1,2}$ and  S.-L.\ Drechsler$^{1}$}
 \address{ $^{1}$Institut f\"ur Festk\"orper und Werkstofforschung Dresden e.V.,
 Postfach 270016, D-01171 Dresden, Germany  \\
 $^{2}$Institute of Spectroscopy,  RAS, Troitsk, 142190, Russia }
\date{May 18, 2001}
\maketitle
\begin{abstract}
The strong coupling Schossmann-Schachinger theory for the upper critical field 
is considered methodically from the user point of view. We check the
accuracy of  approximate formulas for the upper critical field. 
In particular, we explain  in detail as a recently proposed  convenient 
expression (Shulga et al. cond-mat/0103154) can be justified. The 
connection of $H_{c2}(0)$ with the critical temperature $T_c$ is considered 
and the similarity with the Allen-Dynes $T_c$-formula is shown
explicitly.
%IMRAD
\end{abstract}
}
\narrowtext

\section{Introduction}

The standard isotropic single band   (ISB) Eliashberg  model \cite{carbotte90}
is the most developed {\it part} of the  modern  theory of superconductivity.
It describes  quantitatively  the renormalization of the physical properties 
of metals  due to the electron-phonon interaction.
The  input material   parameters are the density of states at $E_F$, $N(0)$,
 the bare Fermi velocity $v_F$, the  impurity scattering rate $\gamma_{imp}$,
 the  paramagnetic impurity scattering rate $\gamma_{m}$, the Coulomb
pseudopotential $\mu^{*}(\approx 0.1)$, and the electron-phonon 
spectral (Eliashberg) function $\alpha^2F(\Omega)$. Thereby the validity of 
the Born approximation is assumed, i.e.\ the system is far from the onset 
of localization.

These days  the material parameters $N(0)$, the averaged Fermi velocity, 
and the spectral function
become available from microscopic
calculations just very soon after the discovery of a new superconducting 
compound. 
The standard procedures of  analysis of many physical properties of metals
include these  results as input parameters. For example, 
the density of states  $N(0)$ is important for the description  of the 
 specific heat data.  Similarly, 
the calculated plasma frequency
 is used for the interpretation of the 
optical and  transport properties. 
As a rule, the forthcoming optical measurements do confirm its value later on. 
 At the same time, the  upper critical field $H_{c2}(0)$, the fundamental 
quantity of a type-II superconductors is usually analyzed by experimentalists
in the framework of  the phenomenological Ginzburg-Landau (GL) theory
which states     
$H_{c2}(0)=\Phi_0/2\pi\xi^2$.
Here  $\Phi_0$ is the  flux quantum and $\xi$ is the GL-coherence length, 
the material parameter of the GL theory.   

Within  the Eliashberg  ISB model
$H_{c2}(T)$ can be computed using the Schossmann-Schachinger
self-consistent theory \cite{shoss86}.   This theory treats  $H_{c2}(T)$ 
as the field  dependence of the critical temperature $T_c(H)$.
At  $H=0$ the linear  $H_{c2}(T)$-problem  is reduced to the  equations
for $T_c$.  From a numerical point of view the eigenvalue  problems for
$T_c$ and $H_{c2}(T)$   are identical and can be  computed by
the same code.   In practice, however, the 
linearized equations for   $T_c$ are frequently 
not exploited even by teams
who perform  tedious numerical band structure or  Eliashberg-theory
calculations. Instead the  well-known phenomenological 
Allen-Dynes expression for  $T_c$ is widely used. In our opinion a similar 
approach could be applied also to  $H_{c2}(0)$ calculating it 
from a simple formula.

Here, for the sake of clarity and simplicity the
Pauli limiting, unimportant in many cases of practical interest, is ignored.
In the present paper we examine five simple expressions
for  $H_{c2}(0)$  by comparison with   the numerical solutions of 
Eqs.\ \ref{ss1}-\ref{ss5}.
% Eliashberg-theory 
One of these formulas valid 
in the clean limit is presented in the present 
 work for the first time. 
% we restricted ourselves  to the consideration 
%of the  clean limit  $H_{c2}(0)$ and neglected Pauli limiting.  
It is shown that at least three  formulas fit reasonably well 
the  numerical values of  $H_{c2}(0)$ in a wide range of coupling strengths
$\lambda$.
 The deviations do not exceed 12\%.  
 Extensions to cases of arbitrary impurity content 
 are considered as well.
 An application example of the 
approximate phenomenological formula is provided.

%In conclusion, the approximate 
%Allen-Dynes expression for  $T_c$ is widely used instead
%of the solution of the corresponding numerical eigenvalue  problem. 
%Similar,   the low temperature value of the upper critical  field  $H_{c2}(0)$
%of clean isotropic metals can be calculated using 
%simple related  formulas.

\section{General equations and  approximate semi-analytical formulas}
\label{appfor}
To take into account fully retardation and strong coupling effects 
in calculating $H_{c2}$, we have solved numerically the 
 linearized 
 %Schossmann-Schachinger 
 equations proposed in 
 Ref.\onlinecite{shoss86}  
\begin{eqnarray}
\tilde{\omega}(n) & = & \omega_n+\frac{\gamma_{imp}{\rm sgn}\omega_n}{2}+
\pi T \sum_{m}  \lambda(m-n)\mbox{sgn}\omega_m,
 \label{ss1} \\
\tilde{\Delta}(n) & = &\pi T \sum_{m}\frac{[\lambda(m-n) - \mu^{*}]
\tilde{\Delta}(m)}{ \chi^{-1}(m)-\gamma_{imp}/2},
 \label{ss2}\\
\chi(n) & = &\frac{2}{\sqrt{\beta}}\int_{0}^{\infty } dq e^{-q^2}
{\rm atan}\left [ \frac{q\sqrt{\beta}}{|\tilde{\omega}(n)|}\right ],
  \label{ss3}\\
\beta & = & \hbar eH_{c2}v_{F}^2/2c,  \label{ss4} \\
\lambda(n) & = & \int_0^{\infty} d\omega \omega\alpha^2F(\omega)/
[\omega^2+(2\pi Tn)^2]. \label{ss5}
\end{eqnarray}
 Here $\omega_n$=$\pi T (2n+1)$ are the Matsubara
frequencies, $e$ and $\tilde{\Delta}(n)$ denote the  electron charge 
 and the superconducting order parameter. $\hbar$ and $c$ are 
 the Plank's constant and  the speed of light.  

From the numerical point of view  the equations of the 
Werthamer-Helfand-Hohenberg (WHH) 
 \cite{werthamer66} and the 
 Schossmann-Schachinger 
 theory have the same order of complexity.
At the same time the WHH equations allow full analytical solutions,
both in the clean and  dirty limits.
 In particular, the clean limit
 WHH orbital upper critical field for  
 the ISB model reads
\begin{equation}
H_{c2}^{{\rm WHH}}(0) =  \frac{k_B^2T_c^2\pi^2c e^{2-\gamma}}{2\hbar ev_F^2},
\label{whh1}
\end{equation}
where $k_B$ and   $\gamma$=0.577  denote
 the Boltzmann's   
 and the Euler's constants.
 In practical units  this formula  can be rewritten as 
\begin{equation}
 H_{c2}^{{\rm WHH}}(0)[{\rm Tesla}]=\frac{0.0231T_c^2[{\rm K}]}
 {v_F^2[{\rm 10^7 cm/s}]}
\approx \frac{0.02T_c^2[{\rm K}]}{v_F^2[{\rm 10^7 cm/s}]}.
\label{whh2}
\end{equation}
Within  the BCS theory, where  2$\Delta\equiv 3.53T_c$,
the formula 
\begin{equation}
H_{c2}^{Sung}(0)=\frac{13.3c\Delta_0^2}{2e\hbar v_F^2},
\label{sungbcs2}
\end{equation} 
is identical to the basic Eq.\ (\ref{whh1}).
It was used by Sung \cite{sung69} for an estimation of $v_F$ in Nb.

McMillan and Werthamer \cite{mcm67} were the first who introduced
%Following to Eliashberg\cite{eli60}
the first order strong
 coupling corrections ($Z(0)= 1+\lambda$) 
 \begin{eqnarray}
&  & H_{c2}^{{\rm McM-WHH}}(T)  =
 H_{c2}^{{\rm WHH}}(T)(1+\lambda)^2, \nonumber \\
&  & H_{c2}^{{\rm McM-WHH}}(0)=
\frac{k_B^2T_c^2\pi^2c e^{2-\gamma}}{2\hbar e[v_F/(1+\lambda)]^2}
=\frac{k_B^2T_c^2\pi^2c e^{2-\gamma}}{2\hbar e{v_F^*}^2}.
\label{mcwhh1}
\end{eqnarray}
They conclude,  that ``this produces a substantial overall increase in $H_{c2}$
from that which would be obtained using the bare band Fermi velocity, but 
being relatively temperature-independent the shift cancels out of $h(t)$.''
These days Eq.\ (\ref{mcwhh1}) is often regarded as a BCS formula since the 
``bare'' Fermi 
velocity has to be replaced by 
 the   ``dressed'' Fermi velocity 
 $v_F^*=v_F/(1+\lambda)$ in Eq.\ \ref{whh1}.
Similarly, the renormalized  form of Sung's formula (\ref{sungbcs2}) reads
\begin{equation}
H_{c2}^{McMS}(0)=\frac{13.3c\Delta_0^2}{2e\hbar {v_F^*}^2}.
\label{sungbcs3}
\end{equation} 
For the sake of completeness, we note that in the presence of 
magnetic pair-breaking 
 the considered renormalization of $v_F$ is different \cite{shulga02}.

Since there are no analytical solution of Eqs.\ (\ref{ss1}-\ref{ss5}), 
approximate expressions should be used  
to estimate $H_{c2}(0)$. 
The parameters in such formulas are determined 
by the fit of numerical data.  
The first expression \cite{carbotte90} reads
\begin{equation}
H^A_{c2} (0)=  
\frac{T_c^2(1+\lambda)^2\pi^2c e^{2-\gamma}}{2\hbar ev_F^2}
\left[1+1.44\left(\frac{T_c}{\omega_{\ln}}
\right)\right],
\label{carb11}
\end{equation}
where $\omega_{\ln}$ is the characteristic phonon frequency introduced 
by Kirzhnits {et al.} \cite{kir73} and used later on in 
Ref.\ \onlinecite{allen75} by Allen and Dynes. 
 A similar $H_{c2}(0)$ expression was proposed 
 by Masharov \cite{masharov74} who obtained however different
  values for the numerical coefficients.

Our {\it factorable } formula, first presented here,  
\begin{equation}
H^C_{c2} (0)\approx 
\frac{k_B^2T_c^2\pi^2c e^{2-\gamma}(1+\lambda)^{2.2}}{2\hbar e{v_F}^2},
\label{our331}
\end{equation}
is suitable for applications together with other  factorable 
expressions.  In practical units it reads
\begin{equation}
H^C_{c2} (0){\rm [Tesla]}\approx\frac{0.0231T_c^2[{\rm K}]
(1+\lambda)^{2.2}}{v_F^2[{\rm 10^7 cm/s}]}.
\label{our33}
\end{equation}
Note that the choice of the coefficients in Eq.\ (\ref{our33}) 
is not unique. For example,  the following expression 
\begin{equation}
H^B_{c2} (0){\rm [Tesla]}\approx 
\frac{0.02T_c^2[{\rm K}](1+\lambda)^{2.4}}{v_F^2[{\rm 10^7 cm/s}]},
\label{our55}
\end{equation}
demonstrates even better agreement with the numerical data. 
Eq.\ \ref{our55} is especially  designed for the Q-check 
\cite{shulga01} in the spirit of the
Langmann's BCS Eq.\ 24 in Ref.\onlinecite{langmann91}.
 
\section{Results and Discussion}

\begin{figure}[t]
\vspace{-0.6cm}
\hspace{-1.3cm}\psfig{figure=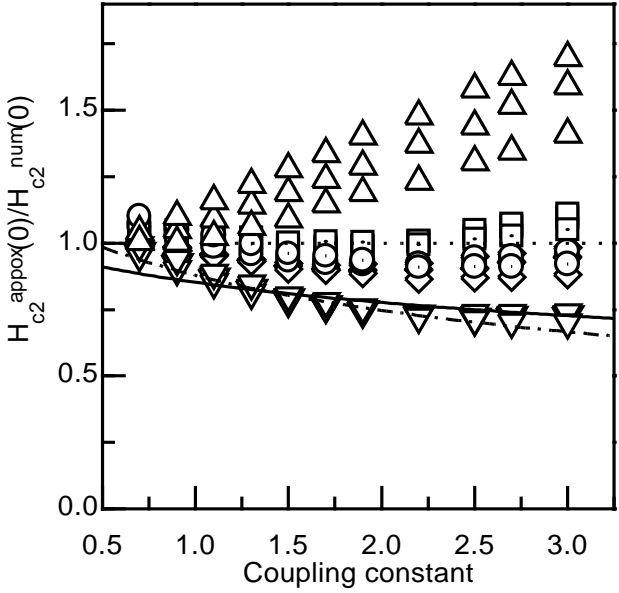,width=8.0cm,height=7.9cm}
\vspace{1.0cm}
\caption{ The upper critical  field,  normalized on the corresponding 
 ``exact''  Schossmann-Schachinger numerical values, {\it vs.} 
 coupling constant $\lambda$.
   Approximate {\it basic} WHH formula 
($\bigtriangledown$, Eq.\ \ref{mcwhh1}),renormalized Sung formula 
 ($\bigtriangleup$, Eq.\ \ref{sungbcs3}), Carbotte formula 
 ($\Diamond$, Eq.\ \ref{carb11}), and our formulas 
 ($\bigcirc$, Eq.\ \ref{our55} and $\Box$, Eq.\ \ref{our33}) were indicated.
 The fitted functions $(1+\lambda)^{-0.231}$ and 
 $0.02(1+\lambda)^{-0.4}/0.0231$ were shown by solid and dash-dotted lines.
 Identical symbols for the same $\lambda$ correspond to different 
 standard values of the 
 Coulomb pseudopotential $mu^*$=0.06, 0.1, and 0.14, respectively.}
 \label{fig1}
 \end{figure}

To check the accuracy of Eqs.\ (\ref{mcwhh1}-\ref{our55}), 
we solved  the Eq.\ (\ref{ss1}-\ref{ss5}) 
employing an Einstein spectral function  with various 
coupling constants $\lambda=0.7-2.9$ and standard values for the 
Coulomb pseudopotential  $\mu^*$.  Then the   
$H_{c2}(0)$-values  
from the approximate formulas 
(\ref{mcwhh1}-\ref{our55}) were normalized  by  the "exact" numerical values.
 The result is shown in  Fig. \ref{fig1}.  If the symbol occurs below the unity line,
the corresponding approximate expression underestimates   
 $H_{c2}(0)$  and vice versa.  
The phenomenological
 formulas  Eqs.\ (\ref{carb11}-\ref{our55}) fit well  the numerical data. 
The deviations from  unity do not exceed 12\% 
in a wide range of coupling strengths.   The formulas Eqs.\ 
(\ref{mcwhh1}-\ref{sungbcs3}) being identical within BCS theory
deviates differently and exhibit relatively  lower accuracy.
%\subsection{Discussion}
 
   A detailed discussion and  collections of the  BCS 
upper critical field formulas   are   given for instance
 in Ref. \onlinecite{fischer82}.
   Here we discuss only the structure  of weak coupling formulas 
Eqs.\ (\ref{mcwhh1}-\ref{sungbcs3}) with respect to the general 
 Bergman-Rainer \cite{rainer74} ansatz
\begin{equation}
H_{c2}(0)=H_{c2}^{BCS}(0)f(\alpha^2F(\Omega), \mu^*, T_c).
\label{ansatz1}
\end{equation}
For the reason of dimensionality $H_{c2}^{BCS}(0)$ is proportional to the 
square of some typical energy. It could be  
$T_c$, the gap $\Delta_0$, the Debye 
energy  $\omega_D$, or the
average phonon frequency $\omega_{ln}$,  and there is a 
total equivalence between them within the BCS  theory due to 
its universal character.
Within the Eliashberg theory  
the situation is different. 
Here Eq.\ (\ref{mcwhh1})
is more preferable than the renormalized Sung's formula (\ref{sungbcs3}) 
(see Fig.\ \ref{fig1}). In fact, at given $\lambda$ 
the $H_{c2}^{McM-WHH}(0)$ points accumulate at nearly  
the same position for different values of $\mu^*$ 
in comparison 
with the
%up-triangles indicating 
$H_{c2}^{McMS}(0)$ data which spread out.  This accumulation
 is a clear advantage 
of the basic  Eq.\ (\ref{mcwhh1}).  A similar choice in favor of 
Eq.\ (\ref{mcwhh1}) was made by  Carbotte  
\cite{carbotte90} without comments. We believe that the weak
sensitivity of the ratio   $H_{c2}(0)/T_c^2$ 
to details of  $\alpha^2F(\omega)$ and the actual value of $\mu^*$ is
due to the fact that $H_{c2}(0)$ and $T_c$ are the solution of the 
same linearized system of Eqs.\ (\ref{ss1}-\ref{ss5}) \cite{shulga98}.
  
   In Sec.\  \ref{appfor} various $H_{c2}(0)$ formulas are  presented in 
different styles.   The first group consists of  ``textbook-like'' 
expressions(\ref{whh1},\ref{mcwhh1},\ref{carb11},\ref{our331}) which
contain  explicitly  fundamental physical and mathematical constants.
Eqs.\ \ref{whh2},\ref{our33},\ref{our55}  written in practical 
units are preferred by insiders working in the field. 
Sung's formulas (\ref{sungbcs2},\ref{sungbcs3})  occupy an intermediate
position.  In our opinion the complete ``textbook-like'' presentation 
is necessary when a formula is derived for the first time from a general
theory without simplifying additional assumptions.  Otherwise, all  
styles are equivalent. In the present context a proof means
a comparison of values given by a formula and a related
computer code.

In view of a reasonable accuracy the question might arise:
``Should  we truncate higher digits in the prefactor $B_1=0.0231$
of Eq.\ \ref{whh2} to $B_2=0.02$ before substitution 
into Eq.\ \ref{ansatz1} or not?" because
the  Fermi velocity is not  constant across the  Fermi surface, and it
is usually known  with some uncertainty.  The  alternative possibilities 
led to Eqs.\ (\ref{our331},\ref{our33}) and  Eq.\ (\ref{our55}).

  The  Carbotte formula (\ref{carb11}) also derived from 
  Eqs.\ (\ref{ss1}-\ref{ss5}) 
through a series of  simplifying assumptions,  includes  second order 
strong coupling corrections in terms of the small parameter $T_c/\omega_{ln}$.
%It  resembles various formulas containing strong coupling correction to the
%BCS expressions. 
The leading  linear term 1.44$T_c/\omega_{\ln}$
 is introduced on a pure phenomenologically  basis 
  to get the best fit of numerical data (see remark below Eq.\ (7.15) 
in Ref. \onlinecite{carbotte90}). 
 The Carbotte formula contains the additional parameter  $T_c/\omega_{\ln}$
in comparison with ``bare'' BCS one and is not factorable. 

For the overwhelming majority of ISB superconductors the  Allen-Dynes formula
\cite{allen75}
\begin{equation}
k_B T_c = \frac{\hbar \omega_{ln}f_1 f_2 }{1.2} \exp \Bigg\{
- \frac{ 1.04 (1+\lambda) }{\lambda - \mu^{*} (1+0.62\lambda) }
\Bigg\}
\quad ,
\label{eq5}
\end{equation}
 describes $T_c$ within  an accuracy of about 5\%. 
In a rigorous sense, it  can  be reduced  to the BCS expression  
\begin{equation}
k_BT_c=1.13\hbar\omega_c\exp\left(-\frac{1+\lambda}{\lambda-\mu^*}\right),
\label{bcstc}
\end{equation}
at $\lambda-\mu^* \rightarrow 0$ only, when $T_c \rightarrow$0. The  
 coupling constant  $\lambda$, the Coulomb pseudopotential $\mu^*$, and 
the characteristic boson energy $\omega_{c}=\omega_{ln}/(1.13*1.2)$ are the
 parameters which enter  Eqs.\ \ref{eq5}, \ref{bcstc}.
Despite the  formal ``contradiction'' to the standard BCS  Eq.\ (\ref{bcstc}),
the Allen-Dynes formula is widely used, especially for qualitative discussions
where possible slight uncertainties may be ignored.
Our formulas Eqs.\ (\ref{our331}-\ref{our55}) were designed in the same manner
and can be treated as the  counter part of the ``Allen-Dynes formulas'' 
for the clean limit $H_{c2}(0)$
of isotropic single band metals. 

$H_{c2}(0)$ does not depend on  $N(0)$ and in view of 
Eq.\ (\ref{ss4}), $H_{c2}(0) \propto 1/v_F^2$. 
As discussed above,  
the relation  $H_{c2}(0)\propto T_c^2$ captures with  high accuracy 
the dependences on $\mu^*$    and on the shape of the spectral function. The remaining 
material parameter $\lambda$  measures 
the overall strength  of the spectral function $\alpha^2F(\omega)$. Since 
$\lambda$ enters the basic BCS formula as the mass renormalization
 (1+ $\lambda$) it is natural at first to make  for 
 the correction function entering Eq.\ 
(\ref{ansatz1})  the following ansatz
\begin{equation}
f=(1+\lambda)^A. 
\end{equation} 
  The {\it chi-by-eye}  \cite{numrec} fit is shown in Fig.\ \ref{fig1} for 
the  prefactor values  $B_1$=0.0231 and $B_2$=0.02. 
The round-off  before  the fit leads to $A_2\approx$0.4   in comparison 
with $A_1$=0.23 obtained for BCS  $B_2\approx$0.0231. 
The additional calculations using various spectral functions and Coulomb 
pseudopotentials 
$\mu^*$   yield a deviation  of $\delta A\approx$0.08. Thus we adopt 
$A_1=$0.2 and $A_2$=0.4.  
Notice that an attempt to renormalize the Fermi velocities by the 
$(1+\lambda)^{1+A/2}\approx (1+(1+A/2)\lambda)$ and the reciprocal 
BCS-coupling 
constant as 
$ 1.04 (1+\lambda) /[\lambda - \mu^{*} (1+0.62\lambda)]$ are  
in a rigorous sense incorrect, since the meaning of ``dressing'' and mass 
renormalization correspond to {\it linear} 
relations. 
  
\begin{figure}
\vspace{-0.6cm}
\hspace{-1.3cm}\psfig{figure=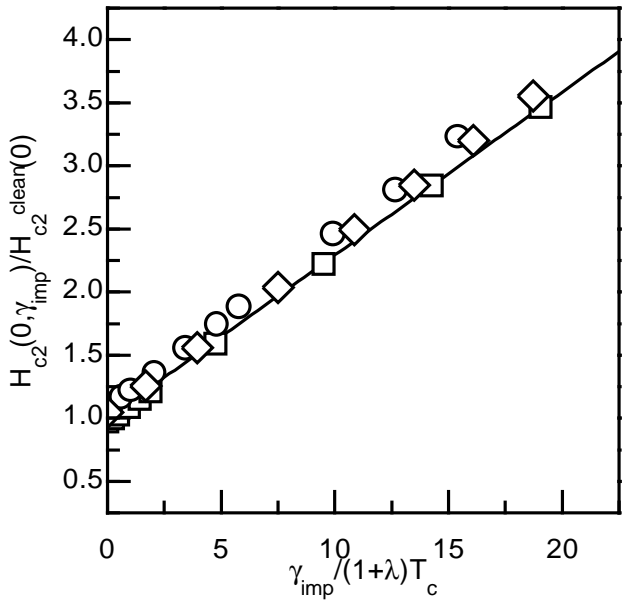,width=8.0cm,height=7.9cm}
\vspace{1.5cm}
\caption{ The calculated  upper critical  field  
for $\mu*$=0.13 and $\lambda=$1 ($\Box$),
2 ($\circ$), 3 ($\Diamond$)
(Eqs.\ \ref{ss1}-\ref{ss5}) normalized on its approximate clean limit 
value (Eq.\ \ref{our33})  
%$H_{c2}(0,\gamma_{imp})/H_{c2}(0,0)$  
$vs$ normalized 
impurity scattering rate $\gamma_{imp}/T_c(1+\lambda)$.
%  $\gamma^*_{imp}/T_c$.
Solid line indicates  BCS approximation (Eq.\ \ref{dirty}).}
 \label{clcrit}
 \end{figure}

%\subsection{The general case of arbitrary impurity scattering rate}
Formally speaking, the discussed above approximate formulas should be 
supplemented by a  clean limit criterion. We remind the reader that there 
is no natural criterion for $H_c2$ in the weak coupling BCS regime. Instead
 the general case of arbitrary impurity scattering rate can be described 
 approximately by a sum of two terms \cite{fischer82}.
It is shown \cite{carbotte90} that the second order corrections 
with an accuracy of about 6\%
coincide in the clean and extreme dirty limits. 
Hence, one expects that  $H_{c2}(0,\gamma_{imp})/H^{clean}_{c2}(0)$,
  the upper critical field 
 at finite $\gamma_{imp}$ normalized on its 
clean limit value, 
can be approximated by the BCS formula.    We 
solved 
Eqs.\ \ref{ss1}-\ref{ss5} for  
Einstein spectra.
The results are shown in Fig.\ \ref{clcrit}. For comparison 
the ratio of the BCS clean and dirty limit expressions
\begin{equation}
\frac{H_{c2}(0,\gamma_{imp})}{H_{c2}^{clean}(0,0)}
\approx 1+\frac
{3\gamma_{imp}}{\pi e^2T_c (1+\lambda)}= 1+\frac
{0.13\gamma_{imp}}{T_c(1+\lambda)},
\label{dirty}
\end{equation}
is depicted too.  
One realizes that this simple BCS-type relationship holds in the 
moderate strongly coupled case, too.

To summarize, for nearly isotropic single band metals
the upper critical field at low temperature can be 
estimated with high accuracy by rather simple semi-analytical formulas
Eqs.\ (\ref{carb11}-\ref{our55}) and Eq.\ (\ref{dirty}).

\acknowledgements
The authors thank H.\ Eschrig, W.\ Weber, 
W.E.\ Pickett, H.\ Rosner, A.A.\ Golubov, O.V.\ Dolgov, and E.\ Schachinger
 for  discussions.
Support from the DFG and the SFB 463 is gratefully acknowledged.

\appendix
\section{ An instructive example from  M\lowercase{g}B$_2$}

 Recently Manske 
 {\it et al.}  \cite{manske01} suggested that  
  experimental data for the novel superconductor MgB$_2$
   which we interpreted in terms of a multi-component gap 
  approach \cite{shulga01}, 
  can be also described within the standard isotropic 
  single-band Eliashberg model using Eq.\ (\ref{carb11}). They reported
  2$\Delta_0 / k_B T_c \approx 4.1$ and  
  H$_{c2}(0)$= 14 T in seemingly good agreement with 
  experimental data for $H_{c2}(0)$. The other reported 
 quantities  are
 $\omega_{ln}$=30 meV, $T_c$=30 K.
 Using their input (material) parameters  presented in that  paper
 we checked their calculations and  find  that their model gives actually
 $\omega_{ln}$=53 meV, $T_c$=26 K, 2$\Delta_0/ k_B T_c = $3.8, and
   $H_{c2}(T=0)$ = 1.2 T in sharp contrast with their set mentioned above.  
   Thus, at first, we confirm our earlier result 
 \cite{shulga01}  that the standard ISB model
 can {\it not} be applied to MgB$_2$. At second, we note that 
 the use of formulas 
 written in practical units might be helpful to avoid 
 such incorrect estimations.

\end{document}